\input harvmac

\def\vp{{\bf p}}

\def \four{{\textstyle {1\ov 4}}}
 \def \third { \textstyle {1\ov 3}} 

\def \ep{\epsilon}
\def\D{\Delta}

\def \x {\xi}

\def \del {\partial}

\def \const {{\rm const}}
\def \ha{{\textstyle{1\over 2}}}

\def \D {\Delta}
\def \a {\alpha}

\def \chi {\chi}

\def \m {\mu}
\def \n {\nu}
\def \vp {\varphi }

\def \t {\tau}
\def \td {\tilde }
\def \d {\delta}

\def \inv {^{-1}}
\def \ov {\over }
\def \four{{\textstyle{1\over 4}}}

\def \X {{\bar X}}

\def \t {\theta} 

\def \D {{\cal D}} 
 
 \def \STr { {\rm STr}}
 
 \def \d {\delta} 
\def \L {{\cal L}}
\def \G {{\cal G}}

\def \lr { \lref}
\def\np {{  Nucl. Phys. }}
\def \pl {{  Phys. Lett. }}
\def \mpl {{ Mod. Phys. Lett. }}
\def \prl {{  Phys. Rev. Lett. }}

\baselineskip8pt
\Title{
\vbox
{\baselineskip 6pt{\hbox{CERN-TH/97-09}}{\hbox
{Imperial/TP/96-97/19}}{\hbox{hep-th/9701125}} {\hbox{revised
  }}} }
{\vbox{\centerline { On non-abelian  generalisation of 
Born-Infeld action 
 }
\vskip4pt
 \centerline {    in string theory     }
}}
\vskip -20 true pt
\medskip
\medskip
\centerline{ A.A. Tseytlin\footnote{$^{\star}$}{\baselineskip8pt
e-mail address: tseytlin@ic.ac.uk}\footnote{$^{\dagger}$}{\baselineskip8pt
Also at Lebedev  Physics
Institute, Moscow.} }

\smallskip\smallskip
\centerline {\it   Theory Division, CERN, CH-1211  Geneve 23,
Switzerland}
\smallskip\smallskip
\centerline {\it  and  }
\smallskip\smallskip
\centerline {\it Blackett Laboratory, 
Imperial College,  London,  SW7 2BZ, U.K. }

\bigskip\bigskip
\centerline {\bf Abstract}
\medskip
\baselineskip10pt
\noindent
We show  that the  part of the tree-level
open string  effective action  for the non-abelian vector field
which depends on the field strength but 
not on its  covariant derivatives, 
is given by the  symmetrised trace 
of the direct non-abelian generalisation of the 
Born-Infeld invariant.
We comment on possible applications to D-brane 
dynamics.

\Date {January 1997}

\noblackbox
\baselineskip 14pt plus 2pt minus 2pt
\lr\ft {E.S. Fradkin and A.A. Tseytlin, \pl B163 (1985) 123.   }
\lr\tse {A.A. Tseytlin, \np B276 (1986) 391; (E) B291 (1987) 876.}
\lr\ts {A.A. Tseytlin, \pl B202 (1988) 81. }
\lr \ant {O.D. Andreev and A.A. Tseytlin, \pl B207 (1988) 157. }
\lr \antt {O.D. Andreev and A.A. Tseytlin, \np B311 (1988/89) 205;
 \pl B207 (1988) 157. }
\lr \mrt {R.R. Metsaev, M.A. Rahmanov and A.A. Tseytlin, 
\pl B193 (1987) 207. }
\lr \grow{ D. Gross  and E. Witten, \np B277 (1986) 1. }
\lr \callan {A.A. Abouelsaood, C.G. Callan, C.R. Nappi and S.A. Yost,
\np B280 (1987) 599. }
\lr \berg { E. Bergshoeff, E. Sezgin, C.N. Pope and P.K. Townsend,
\pl B188 (1987) 70.}
\lr \bergg { E. Bergshoeff, M. Rakowski and E. Sezgin, 
\pl B185 (1987) 371.}
\lr\frat {E.S. Fradkin and A.A. Tseytlin, \pl B160 (1985) 69. 
  }
\lr \leii{
R.G. Leigh, \mpl A4 (1989) 2767. }
\lr \lei{J. Dai, R.G. Leigh and J. Polchinski, 
\mpl A4 (1989) 2073.}
\lr \pol { J. Polchinski, hep-th/9611050. }
\lr  \witt { E. Witten, \np B443 (1995) 85, hep-th/9510135. } 
\lr \bach { C. Bachas \pl B374 (1996) 37, hep-th/9511043. }
\lr \tst { A.A. Tseytlin,\np B469 (1996) 51,  hep-th/9602064. }
\lr \tseyy { Tseytlin  Int.J.Mod.Phys.A4:1257,1989.} 
\lr \scherk {A. Neveu and J.  Scherk, \np B36 (1972) 155;
J.  Scherk  and J.H. Schwarz, \np B81 (1974) 118. }

\lr \napp {P.C. Argyres and C.R. Nappi, \np B330 (1990) 151. }
\lr \kita {Y.  Kitazawa, \np  B289 (1987) 599.} 
\lr \prev { T. Hagiwara, J. Phys. A14 (1981) 3059.}
\lr \dorn {H. Dorn, hep-th/9612120; 
H. Dorn and H.-J. Otto, hep-th/9603186.}
\lr\ishi{
N. Ishibashi, H. Kawai, Y. Kitazawa and A. Tsuchiya, hep-th/9612115.}

\lr \anmpl {O.D. Andreev and A.A. Tseytlin, \mpl   A3 (1988) 1349.}

  \lr \fri{ C. Lovelace, \np B273 (1986) 413; 
  B.E. Fridling and A. Jevicki, \pl B174 (1986) 75.}
  
 \lr \hamada { K. Hamada, hep-th/9612234.}
  \lr \shif { M.A. Shifman, \np B173 (1980) 189.}
   \lr\nev {G.L. Gervais and A. Neveu, \np B163 (1980) 189; H. Dorn,
 Fortschr. d.  Phys. 34 (1986) 11.} 
 \lr\schw{M. Aganagic, C. Popescu and J.H. Schwarz, hep-th/9612080.}
 
 \lr\brow{H. Leutwyler, \np B179 (1981) 129;
 L.S. Brown and W.I. Weisberger, \np B157 (1979) 285.}
 \lr \doug {M. Douglas, hep-th/9512077.}
 \lr \gre{M.B. Green and N. Seiberg, \np B299 (1988) 559.}
 
 \lr \ishi {N. Ishibashi, H. Kawai, Y. Kitazawa and A. Tsuchiya, 
 hep-th/9612115.}
 
 \lr\lii{M. Li, hep-th/9612222.}
 
 \lr \grg{ M.B. Green and M. Gutperle, hep-th/9612127.} 
\lr \periw{ V. Periwal, hep-th/9611103.   } 

 \lr\banks{
T. Banks, W. Fischler, S.H. Shenker and L. Susskind, 
   hep-th/9610043.}

\lr\doug{M. Douglas, hep-th/9512077.} 
\lr\kab{M.R. Douglas, D. Kabat, P. Pouliot and  S.H. Shenker,
  hep-th/9608024. }
 
 \lr\tayl{W. Taylor, hep-th/9611042.}
\lr \gano{O.J. Ganor, S. Ramgoolam and W. Taylor, 
  hep-th/9611202.}
  \lr\motl{L. Motl,   hep-th/9701025.}
\lr\egu{T. Eguchi and H. Kawai, \prl 48 (1982) 1063;
D. Gross and Y. Kitazawa, \np B206 (1982) 440; 
S. Das and S. Wadia, \pl B117 (1982) 228.}
 
 \lr \schm{C. Schmidhuber, \np B467  (1996) 146. }
 
 \lr\bankk{ T. Banks, N. Seiberg and S. Shenker, hep-th/9612157.}
 
 \lr \perw{V. Periwal, hep-th/9612215.   } 
 \lr\bas{T. Banks and L. Susskind, hep-th/9511194.}
 \lr\nish{H. Nishino and E. Sezgin, hep-th/9607185.}
 \lr\john{J.H. Schwarz, \pl B360 (1995)13, hep-th/9508143.}

\lr\towns{P.K. Townsend, \pl B373 (1996) 68,   hep-th/9512062.}
 \lr\dull{M.J. Duff and J.X. Lu, \np  B390 (1993) 276.}
 
\lr\gil{G. Lifschytz, hep-th/9612223.}
\lr\gilma{G. Lifschytz and S.D. Mathur, hep-th/9612087.}
\lr\ahber{O. Aharony and M. Berkooz, hep-th/9611215.}
\lr\make{I. Chepelev, Y. Makeenko and K. Zarembo, hep-th/9701151.}
\lr\mye{J.C. Breckenridge, G. Michaud and R.C. Myers, hep-th/9611174.}
\lr\rust{J. Russo and A.A. Tseytlin, hep-th/9611047.}
\lr \liff{G. Lifschytz, hep-th/9610125.}
\lr\tsey{A.A. Tseytlin, hep-th/9612164.}


Non-locality of string theory (i.e. the presence of a
tower of massive states) implies that   the  low-energy effective
action  for massless modes  is an   infinite power series  
of all orders in 
$\a'$ \scherk.  In particular,  this applies to the tree-level 
Lagrangian $L_{eff}$  for the  gauge vectors  in the open 
bosonic or  type I string theory. 
In the  case of  an abelian Chan-Paton gauge group  all 
terms in the action which depend on the field
strength $F_{mn}$ but not on its derivatives  
sum up  into the Born-Infeld (BI) Lagrangian 
  \refs{\ft,\tse,\callan,\ts,\berg,\mrt} (we use the euclidean
  signature)
\eqn\born{
L_{BI} = c_0 \sqrt {\det(\d_{mn} 
+ T\inv  F_{mn} ) } \ ,  \ \ \ \ \  \  T\inv = 2\pi \a' \ . }
Derivative corrections to this  
action were discussed in \refs{\antt,\anmpl,\kita}.

In  the non-abelian case, the tree-level (disc) 
 effective Lagrangian  in the open  string theory 
   can be represented  as an expansion in powers of the 
 field strength and  its covariant derivatives, 
\eqn\expa{
L_{eff}  = \Tr \big( a_0 F^2 + a_1 FD^2F + a_2 F^4  +  a_3 F^2D^2F + ...
\big)  = 
L (F) +  O(DF) \ , }
where $L(F)$ is the part not containing   covariant derivatives of
  $F_{mn}= \del_m A_n - \del_n A_m - i [A_m,A_n]$
  ($F$ is assumed to be 
 a hermitian 
 matrix with indices in the fundamental representation of the gauge
 algebra).
 Previously,  only  the  terms up to order  $F^4$  in  \expa\ 
 were  completely
determined \refs{\tse,\grow} (there was also
a discussion of
$F^5$ terms in \kita). 
 The question we shall address below is  about the structure of 
 $L$ in \expa, i.e. of  a 
 non-abelian analogue  of the BI action (NBI action for short).

In contrast to the abelian case where the separation  between 
derivative-independent and derivative-dependent terms
 in $L_{eff}(F,\del F)$
is completely unambiguous, 
this is not true
in the non-abelian case.
Since  $[D_m,D_n] F_{kl} =
 [F_{mn},F_{kl}]$   some of the derivative terms may be traded for
 some of non-derivative ones, and vice versa. We shall resolve this ambiguity 
 by assuming that   all $[F,F]$ (`commutator')  terms
should be treated as a part of the  $DF$-dependent terms in $L_{eff}$ and thus 
should {\it not}  be included into  $L(F)$ in \expa. 
The effective Lagrangian will then be dominated by $L(F)$
 under the circumstances 
when the covariant derivatives of $F$ are much smaller than
the powers  $F$. 

Adopting such a definition of $L(F)$ or NBI Lagrangian, we shall 
prove below that, both in the bosonic and the superstring theory, 
 it is given by  the following natural generalisation 
of the  Born-Infeld action \born\
 \eqn\nbi{
 L(F)  =L_{NBI}= c_0 \STr \sqrt {\det(\d_{mn} 
+ T\inv  F_{mn} ) } \  . }
Here $\d_{mn}$ implicitly includes a factor of   
the  unit matrix in internal space,
 the determinant
is computed with respect to the $mn$ indices only,  
and $\STr$ is the symmetrised trace in the fundamental representation,  
 $\STr (A_1 ...A_n) \equiv  {1\ov n!} \Tr(A_1...A_n $ + all 
 permutations).
 This Lagrangian  is thus equal to the same sum of even powers of
  $F_{mn}$  as  appearing 
  in the expansion of BI Lagrangian \born,
   with each factor of field strength     being replaced
   by  a hermitian matrix $F$ and all possible orderings of the matrices 
   included with equal weight.
 The same invariant 
  was  previously  conjectured  to be a  part of a  
non-abelian generalisation 
of BI Lagrangian in \napp, 
 where, however, an additional term
with $\STr$ replaced by the antisymmetrised trace was also 
suggested  to be present.\foot{Some other  ad hoc   generalisations of BI
action to non-abelian case 
 where  considered     in \prev\ but because of their
 different trace structure they cannot appear 
 in the  tree-level open string effective action. }
 The latter is given by the sum of  traces of odd powers of $F$ which 
  always contain a factor of $[F,F]$  (as follows from 
   $F_{mn}=-F_{nm}$) and thus should not be included into NBI Lagrangian 
   according to
 the  definition given above.

 Let us first compare the  $\a'$-expansion  of \nbi\ 
 ($c_1=   \pi^2 \a'^2 c_0 $)
 $$  
 L'_{NBI} = c_0 \STr [\sqrt {\det( \d_{mn} 
+ T\inv  F_{mn})} -I]  $$  $$
= c_1 \STr \big[ F^2_{mn}  -  \ha (2\pi \a')^2
 \big(F^4 - \four (F^2)^2\big) + O(\a'^3) \big] $$
  \eqn\nbe{
 = c_1 \Tr \big[
F^2_{mn} -  {\third} (2\pi\a')^2 \big(F_{mn}F_{rn}F_{ml}F_{rl} +
\ha  F_{mn}F_{rn}F_{rl} F_{ml}  } $$ 
- \  \four F_{mn}F_{mn}F_{rl}F_{rl}
-{\textstyle {1\ov 8}} F_{mn}F_{rl}F_{mn}F_{rl} \big)
  + O(\a'^4) \big] \ ,  $$  
  with  the   known 
 perturbative results.
The two leading orders in $\a'$ in \nbe\ 
 indeed  give  the full  form of the non-abelian 
open superstring effective action to order $O(\a'^3)$  
(all $\a'^2$-terms with covariant derivatives 
 have field redefinition dependent coefficients \tse).
The $F^4$ terms were originally 
found in the $\STr$-form in \grow\ and in the equivalent 
$\Tr$-form in \tse.


As for the bosonic theory, there 
\nbe\ does not represent 
 the full  effective Lagrangian to $\a'^3$-order:
 the bosonic $L_{eff}$  contains  $\a'F^3$ 
 term  \scherk\  and 
 the coefficients of the  $F^4$ invariants are somewhat different 
 from the ones in \nbe\ \tse. However,   it is easy to see 
  that both $F^3$ and 
  the excess of $F^4$ 
  terms are the `commutator' terms, i.e. they 
  can be represented as $ \Tr ( {4\ov 3} i\a' 
  F_{mn} [F_{ml}, F_{nl}] +  
  2\a'^2 \Tr ( F^{mn}F^{rl}[F_{mn},F_{rl}])$
   and thus, 
  according to our definition,
   belong to  the covariant derivative 
  part of  $L_{eff}$ and not to the 
   NBI part. Similar remark applies to the
   $F^5$  terms  \kita\foot{The $F^5$-terms have 
   the coefficients proportional
  to $\zeta(3)$ \kita\  and  
   should rather not appear in any simple NBI
  action.}
  and, in general, to all terms of 
  odd power in   $F$.

Let us now  give the general argument demonstrating   that 
the  covariant derivative independent part 
of the open string effective action is indeed given by the 
 NBI action  \nbi. The starting point is the expression for the 
 generating functional for the 
   vector amplitudes on the disc.
 In the bosonic case  \refs{\ft,\tse}
 \eqn\act { 
 Z(A) = < \Tr P \exp [i\int d \vp \  \dot x^m A_m (x)] >  } $$
 = \int d^D x_0  < \Tr P 
  \exp[i \int d \vp \  \dot \x^m A_m (x_0 + \x)]> \  , $$
 where $ x= x_0 + \x (\vp)$, \ 
 $0 < \vp \leq 2\pi$ and 
 the averaging is done with the free string propagator 
 restricted to the boundary of the disc ($\ep\to + 0$ is a  world-sheet 
  UV regularisation)
 \eqn\avv{ < ... >  =  \int [d\x] \ e^{ - { 1\ov 2} T \int \x G\inv
   \x  }  ... \ , \ \ \ \       
   G(\vp,\vp') = {1\ov \pi}\sum^\infty_{n=1} {e^{-n\ep}
   \ov n} \cos n (\vp-\vp') \ .  }
As explained in \refs{\ts,\antt},  the low-energy effective action 
is given by the renormalised value of \act, computed by expanding in 
powers of $\a'$, \  
$S_{eff} (A) = Z(A(\ep),\ep)$.\foot{The logarithmic 
renormalisation of the `coupling' $A_m$  corresponds to  a 
subtraction of the massless poles in the amplitudes \refs{\fri,\tse,\ts}\
(the field redefinition ambiguity in the effective 
action  corresponds to the 
renormalisation scheme ambiguity in this framework \tse).
In addition, one is to subtract (or absorb into the
renormalisation of the tachyon coupling)
 the 
leading linear divergence. This is equivalent   to a 
 subtraction 
of the 
$SL(2,R)$ M\"obius group  volume factor. Power divergences are 
absent  in the
superstring case where the super-M\"obius  volume is finite \antt.
}

Using the radial gauge $\x^m A_m (x_0 + \x) =0, \ A_m (x_0) =0$
(see, e.g., \shif)   we get the following expansion 
in terms of symmetrised products of covariant derivatives of $F$
 at  $x_0$, 
 \eqn\rad{ 
 \int d\vp\ \dot \x^m A_m (x_0 + \x) =  \int d\vp\ \dot \x^m\big[
 \ha \x^n F_{nm} + 
  {\textstyle {1\ov 3}}\x^n \x^l D_l F_{nm} + 
   {\textstyle {1\ov 8}} \x^n \x^l \x^s D_{(s } D_{l)} F_{nm} + ... \big] \ .  }
Separating  in this way 
the dependence of $Z$ on covariant derivatives  we are led to 
 \eqn\actr { 
 Z(A) =  \int d^D x_0
  \big[ \L (F)  + O( D_{(k} ... D_{l)} F) \big] \ ,  }
  \eqn\ttt{
 \L (F) =  < \Tr P 
  \exp\big[\ha { i} F_{nm}  \int  d \vp \  \dot \x^m \x^n \big] >  \  .  }
 The path integral in  \ttt\    is effectively non-gaussian\foot{It 
 may be re-written as a standard
  1-dimensional  path integral by 
   introducing the auxiliary
   fields to represent the path-ordered exponent as,  e.g.,  in
    \refs{\nev,\antt}. }
     because
 of the  normal ordering of the  $F_{nm}(x_0) (\dot \x^m \x^n)(\vp)$ 
 factors
 which is non-trivial if the matrices $F_{mn}$ do not commute. 
   It may still be possible to
 compute it explicitly. 
In the abelian case the path ordering is trivial  and one finds 
\eqn\yyy{L(F) = c_0 \big[{\det(\d_{mn} 
+ T\inv  F_{mn} ) }\big]^\n\ ,   } \eqn\nen{
\ \ \  \n = - \pi \int^{2\pi}_0 \dot G^2 
= - \big(\sum^\infty_{n=1} e^{-2\ep n} \big)_{\ep \to 0} 
 = -{\textstyle {1\ov 2 \ep }} + \ha  \ , }
so that $\n= \ha$ after the  subraction of the 
M\"obius volume divergence
\ts\ (which is  done effectively  when   using the 
$\zeta$-function prescription \ft). As a result,  
one finds the BI expression
\born.

Since  we defined the $DF$-independent part $L(F)$ of the
effective Lagrangian as not  containing terms with commutators 
of $F$, 
 to determine it   we may treat the matrices $F_{mn}$ in 
 \ttt\ as 
commuting, or, equivalently, symmetrise over all of 
their orderings
in each monomial $F^n$. Then the  path ordering becomes trivial
just as in the case of the abelian gauge group,  
 so that  instead  of \actr\ we get   
\eqn\actre { 
 Z(A) =  \int d^D x_0
  \big[ L (F)  + O( D_{k} ... D_{l} F) \big] \ ,  }
 and 
\eqn\actro { \L(F)\  \to\ 
 L (F)  = < \STr P 
  \exp[\ha i F_{nm} \int  d \vp \  \dot \x^m \x^n ] > }
  $$
   = \STr <  
  \exp [\ha i F_{nm} \int^{2\pi}_0  d \vp \  \dot \x^m \x^n ] >  
   = c_0   \STr \big[  {\det(\d_{mn} 
+ T\inv  F_{mn} ) }\big]^\n  \  .   $$
 Since $\n_{ren}= \ha$, we finish with   the   NBI Lagrangian \nbi. 
 
 This discussion is readily generalised to the 
 superstring case, where  the 
 gauge-invariant expression for the generating functional 
 is given by the following manifestly  1-d supersymmetric expression
 \antt\
 \eqn\acts { 
 Z(A) = < \Tr \hat P \exp [i\int d \hat \vp
  \  \D \hat x^m A_m (\hat x)] > 
  \  . \  }
 Here 
 $\hat x^m = x^m (\vp)   + \theta \psi^m (\vp)  , \ \  d \hat \vp  = d\vp
 d\t , $ $
 \ \D= {\del \ov \del \theta} - \theta  {\del \ov \del \vp}$
 and the supersymmetric path ordering 
 $ \hat P$ is defined by replacing the usual $\Theta$-functions by the 
supersymmetric ones, 
$\hat \Theta (\hat \vp_i, \hat \vp_j) = \Theta (\hat \vp_{ij})
= \Theta (\hat \vp_i - \hat \vp_j)
 + \theta_i \theta_j \delta (\vp_i-\vp_j), \ 
\hat \vp_{ij}\equiv  \vp_i -\vp_j + \t_i\t_j$, 
so that $\D \hat \Theta$ is equal to the 
supersymmetric $\delta$-function
$ \delta (\hat \vp_{ij}) =  (\theta_j   -\theta_i    )  
\delta (\vp_i-\vp_j)$.
The generating  functional \acts\ 
automatically includes the contact terms necessary \gre\  for
maintaining  gauge invariance. Re-written in terms of the
standard path ordering,  it takes the form \antt\
\eqn\ats { 
 Z(A) = < \Tr  P \exp\big( i\int d \vp
  \ \big[\dot x^m A_m (x) - \ha \psi^m\psi^n F_{mn} (x) \big] \big) > 
  \  ,   }
with the $[A_m,A_n]$ term in $F_{mn}$ appearing due to the 
presence of the  contact 
$\theta_i \theta_j \delta (\vp_i-\vp_j)$  terms
in the supersymmetric theta-functions in \acts.
The definition of $<...>$ is  analogous to  \avv\ 
with  $\x G\inv \x \to \x G\inv \x + \psi K\inv \psi $,  
\  where $K$ is the restriction of the fermionic Green's 
function to the boundary of the disc, 
 $$ K(\vp,\vp') =   {1\ov \pi}\sum^\infty_{r=\ha } {e^{-r\ep}
  } \sin r (\vp-\vp') \ . $$
As a result, the superstring 
generalisation of \actr,\ttt\   has $\L(F)$ given by 
\eqn\actrs { 
 \L (F) =  < \Tr P 
  \exp\big[\ha i F_{nm}  \int  d \vp \  
  ( \dot \x^m \x^n  +  \psi^m \psi^n ) \big] >  \  .  }
Dropping the `commutator' terms to define $L(F)$, i.e. symmetrising the trace, we 
get,  as in \actro,\yyy,   
 \eqn\actrss { \L(F)\  \to\ 
 L (F) 
   = \STr <  
  \exp\big[{\ha i F_{nm} \int^{2\pi}_0  d \vp \ (
   \dot \x^m \x^n  +  \psi^m \psi^n )} \big] >   } $$ 
   = c_0   \STr \big[ {\det(\d_{mn} 
+ T\inv  F_{mn} ) }\big]^\n  \  ,  $$
where now 
\eqn\nnn{ \n = - \pi \int^{2\pi}_0  ( \dot G^2  -  K^2) 
=( - \sum^\infty_{n=1} e^{-2\ep n} +  \sum^\infty_{r=\ha }
 e^{-2\ep r})_{\ep \to 0}  = \ha   \ . }
Thus  we again 
 obtain  the NBI Lagrangian \nbi, 
here   in completely unambiguous way
  as the linear divergence in $\n$ present in bosonic case 
    cancels out 
\mrt\ 
(which is  a manifestation of 
 the finiteness of the  volume of the super-M\"obius
group \antt).

To summarise, the NBI action \nbi\ is thus a good approximation  to the effective action
when all products of covariant derivatives of $F$  are small.
Since  $[D,D] F= [F,F]$
 that also means that the 
`commutator'
terms are assumed to be small, 
i.e. the field strength is approximately
abelian.\foot{There is also another  choice for a translationally
invariant 
non-abelian gauge field: $A_m=\const$ \brow.
 It would be interesting to compute 
the  value of the effective action, i.e. the partition function \acts,\ats\
in this case.}
  There may be physically interesting 
  cases  in which  such 
an approximation is 
a useful one. 

There is a  
possible alternative expansion of $L_{eff}$ 
in which one assumes that all {\it symmetrised} covariant derivatives are
small. This  
 does not imply smallness of commutators of $F$. In this case,
as follows from the discussion above (see \actr), the effective  Lagrangian  is 
approximated by $\L(F)$ in \actrs\
(for which, unfortunately, 
   we do not know a closed expression). 
    $\L(F)$ and $L(F)$ are
   the same at $F^4$-order 
   in the superstring case (but not in the bosonic case) but 
   are expected
   to differ at higher orders in $F$.

The Lagrangian $L(F)=L_{NBI}$ \nbi\ is  thus the 
 simplest  and natural  non-abelian generalisation of the BI Lagrangian. It  
  should admit a $D=10$ supersymmetric extension
 generalising the  action  
found in  the abelian case in \schw\ 
 (a supersymmetric  version of 
 $k_0 \Tr F^2  +  k_1\  \STr[F^4 -\four (F^2)^2]$  with the 
symmetrised  trace  is known to exist 
\bergg). One  indirect  attempt 
to find the  supersymmetric NBI action 
could be   to repeat the above analysis using the 
light-cone Green-Schwarz formalism  with 
 the  fermionic partner
of $A_m$ included in the world-sheet action (cf. 
\refs{\frat,\tse,\hamada}).

Let us now comment on  the   application 
 of NBI action \nbi\ to 
the description of D-branes \refs{\lei,\pol}. 
We shall only consider the 
D-brane motion in  a trivial flat background. 
The form of   the 
D-brane  effective 
action   \refs{\leii} 
is essentially determined   
(via T-duality)  
 by  the abelian $D=10$   open 
superstring effective action  
(see \refs{\bach,\tst}).
In the `small acceleration' approximation  it is 
thus  given    
by the BI action for the $D=10$ vector potential 
$A_m  =(A_s, A_a= T X_a)$   reduced to $p+1$ dimensions.
This leads to  the  D-brane action in the static gauge\foot{Here
we switch from the Euclidean 
to the  Minkowski signature and 
use the following notation:
 $m,n=0,1,...,9$;\  $r,s= 0,1,...,p$; \ $a,b= p+1, ..., 9$. 
 The functions $A_s$ and $X_a$ depend only on $x_s=(x_0,...,x_p)$.
 In the case of D-instanton action below the metric will be again
 Euclidean, i.e. $X_0$ will be the Euclidean time direction.}
\eqn\qer{
I_p = T_p \int d^{p+1} x \sqrt {-\det 
(\eta_{mn}  +   T\inv F_{mn})} } $$ = 
 T_p \int d^{p+1} 
  x \sqrt {-\det 
(\eta_{rs}  + \del_r X_a \del_s X_a +   T\inv F_{rs})}\  . $$
In the low-energy or `non-relativistic' approximation, i.e. 
to the leading quadratic order in $F_{mn}$,  this action
  is the same as the 
dimensional reduction  of the $D=10$  $U(1)$ 
Maxwell action for $A_m$  \witt.
As argued in \witt, for a   system of 
$N$ parallel D-branes the fields 
$(A_s, X_a)$ become $U(N)$ matrices and the 
  Maxwell action  is generalised to the 
$D=10$  Yang-Mills action reduced to $p+1$ 
dimensions. 

This action  is, in general,  
corrected by higher-order
terms  which, as in the abelian case, are
determined by   the 
dimensional reduction
of the   open string effective action with the gauge potential 
components replaced by the matrix-valued fields $(A_s, A_a= T X_a)$.
This follows directly   from the  non-abelian 
generalisation of the  partition function approach to the derivation of
D-brane actions  discussed in \tst\ (T-duality  relates 
the Neumann $A_s$ and Dirichlet $A_a$ vertices in the exponent in 
 $Z= < \Tr P \exp \{i\int d\vp [\del_\vp x^s A_s(x) + 
 \del_\bot x^a A_a(x)]\} >$).\foot{For a  discussion 
 of  leading-order  D-brane equations in the non-abelian case in the 
 alternative conformal invariance  approach (used in the abelian case in
 \leii)   
 see \dorn.}
It is natural to  expect that the  most important part of these
corrections is  represented
 by the NBI action \nbi, i.e. by the following generalisation
 of \qer\ 
  \eqn\rrer{
I_p = T_p \int d^{p+1} x \
 \STr \sqrt {-\det 
(\eta_{mn}  +   T\inv F_{mn})}  }
$$ = 
  T_p   \int d^{p+1} x \
 \STr  \big[  
  \sqrt {-\det 
(\eta_{rs}  + D_r X_a (\d_{ab} -i T [X_a,X_b])^{-1} D_s X_b
 +   T\inv F_{rs})}    $$ 
$$ \times  \sqrt {\det (\d_{ab} -i T [X_a,X_b]) } \ \big] \ . $$
Here $\STr$ applies to the products of 
 components of the field strength $F_{mn}$, i.e.
$F_{ab} = -iT^2 [X_a,X_b]\ , 
 \ F_{ra} = T D_r X_a= T(\del_r X_a - i [A_r,X_a])$ and $F_{rs}(A)$.
Expanding in powers of  $[X_a,X_b]$ 
  we  find  (cf. \pol)
  \eqn\ere{
  I_p= T_p   \int d^{p+1} x \ \STr \big[ 
  \sqrt {-\det 
(\eta_{rs}  + D_r X_a  D_s X_a
 +   T\inv F_{rs})} -  \four T^2  ([X_a,X_b])^2 
 + ... \big] \  .   }
 Note that the  first term here gives the full 8-brane action 
 where there is only
 one  $X_a=X_9$.
 
 Like  the  abelian BI action \qer\
with all higher-order $F^n$
terms included which 
grasps some  important features of  D-brane dynamics 
(e.g., a 
relation  between  the existence of limiting  velocity and 
  maximal
field strength \bach) the   NBI action \rrer\
  may also  find some useful applications, 
  provided one understands  
   the regions of  applicability
  of different expansions used.
  
Originating from the open string theory, the above D-brane 
actions 
should be related by $T$-duality  which  transforms  a system
of Dp-branes in the space with one circular dimension
into  D(p+1)-branes wrapped around the dual 
circle.  This  can be demonstrated 
following \refs{\banks,\tayl,\gano,\motl}. 
  One   formal way 
to get the D(p+1)-brane action from the Dp-brane  action  
is to add dependence on an `auxiliary' 
  parameter  $x_c$ and 
replace  the   collective coordinate $X_{c}$ corresponding to the 
compact  dimension transverse to the  Dp-brane   by the
covariant derivative operator $i\del_{c} + A_{c}(x_s, x_{c})$. 
  This effectively accounts for  the presence of 
  the open string winding
  modes, and is also  
 equivalent to `gauging' the translations in this direction
(by embedding the translations into the 
gauge algebra or, going in the opposite $p+1\to p$
direction, by replacing the tensor product of the matrix algebra 
and the algebra of linear differential 
operators on functions of $x_{c}$ by another matrix algebra, 
which is 
possible  in the large $N$ limit).
This  adds  an extra dimension to the world volume and 
trades  one $X_{c}$ for $A_{c}$.

For example, let us consider  the action \rrer\ 
for  $N$ instantons, $p=-1$  (cf. \refs{\witt,\ishi,\grg})
 \eqn\inst{
I_{-1}  = T_{-1}\ 
 \STr  \sqrt {\det (\d_{mn} -i T [X_m,X_n]) } }
 $$
 =  T_{-1} 
 \ \Tr \big[ 1 -  \four T^2  ([X_m,X_n])^2  +  ...\big] \ .   $$
If we assume that the euclidean time direction corresponding to $X_0$  
is compactified,   then 
 the prescription is  to  extend  the gauge algebra 
  by introducing the `dual'
  coordinate $x_0$  so that the $U(N)$ matrices 
 $X_a$ ($a=1,...,9)$  become functions of $x_0$
  and  $X_0$ becomes a matrix-valued linear 
 differential operator, \  
  $ T X_0 = i\del_0 + A_0 (x_0) $\  (the action of $\del_0$ is
  understood in the sense that 
    $[X_0,X_a] = iT\inv D_0
  X_a$, etc.). 
 As  follows  from  \rrer,  
  the   D-instanton action \inst\  then transforms into the 
  (multi) 0-brane action (after $\STr \to \int dx_0\ \STr$ and $x_0 \to
  ix_0$) 
   \eqn\zerr{
I_0 = T_0 \int d x_0 \
 \STr  \big[  
  \sqrt { 
(1-    D_0 X_a (\d_{ab} -i T [X_a,X_b])^{-1} D_0 X_b)} 
  \sqrt {\det (\d_{ab} -i T [X_a,X_b]) }\  \big] .  }
  Related  observations at the
 leading-order level  of reduced Yang-Mills actions 
were made in
\refs{\periw,\ishi,\motl}.
Similarly, one can transform   the  D0-brane action into the 
  D-string action, or into the D2-brane action, etc.

  Expanded in  powers of velocities,  
   the 0-brane  action \zerr\  becomes  (e.g., in the $A_0=0$ gauge)  
  \eqn\zeee{I_0=   T_0 \int d x_0 \ 
 \STr  \big( \sqrt {\G }  \big[ 1 - 
 \ha \G^{ab}(X)  \dot X_a \dot  X_b  + 
 O(\dot X^4)\big]  \big) \ , }
 $$ \G_{ab} \equiv   \d_{ab} -i T [X_a,X_b] \ . $$
The matrix-valued  `metric'   $\G_{ab}$
may have some  `non-commutative geometry'  interpretation.

Let us now discuss  a  
closely related  approach to  
 correspondences between different D-brane
actions, 
 which is  in the spirit of  the  matrix models of 
   \refs{\banks,\ishi,\bankk}.
The aim will be to identify higher D-branes as solutions 
of the instanton model \inst\
and to reproduce their actions by expanding 
\inst\ near the classical background as in \refs{\bankk,\lii}. 
 The $U(N)$ action \inst\ for a system  of   D-instantons
 is convex just like the (euclidean)  abelian BI action.
 As  the leading-order  reduced Yang-Mills action 
 $\sim ([X_m,X_n])^2$,  it  has the absolute minimum at 
 $[X_m,X_n]=0$.\foot{It is interesting to note that 
 in the case of a system of 
 $n$ instantons and $N-n$ anti-instantons  which, 
 if we follow \perw,
 may be 
 described by 
 \inst\ with non-compact $U(n,N-n)$ gauge group,
  the action
 is no longer convex and not 
 even real for arbitrary  $X_a$,  so that  there is a possibility  
 of  non-trivial minima and tachyonic instabilities (cf.
 \refs{\bas,\perw}).   This should be   analogous 
 to the relation between
 the existence of a  maximal field strength in the BI action and the  
 tachyonic instability of  the open string theory in a 
 large vector field background. Here the role of field strength 
 is effectively played by the commutator of coordinates $X_a$.
 This is one of the examples for which the difference between 
 the   NBI instanton action \inst\ and the reduced  Yang-Mills action 
 should become   crucial.}
 The equations of motion following from 
  the  euclidean  NBI action \nbi\
 are solved, in particular, 
  by the covariantly constant fields, $D_k F_{mn}=0$.
  Such configurations are precisely 
  the ones for which  our approximation  based on neglecting  the 
  covariant derivative  
  corrections to the NBI term in \expa\ is {\it justified}. 
 In general, the  `covariantly constant' classical 
  configurations  corresponding to \inst\ satisfy
  (bar will  denote  a classical
solution)
 \eqn\eqe{  [\X_m,[\X_n,\X_k]]=0\ .  }
 The non-trivial {\it vacuum} 
   configurations have  $[\X_m,\X_n]=0$,  with
 some  $\X_m\not=0$. 
 The solutions with $[\X_m,\X_n] = i f_{mn} I$ 
  (where $I$ is a unit matrix)  
 which exist only in the large $N$ limit   correspond to 
 {\it solitonic} BPS
 configurations
 \refs{\banks,\ishi,\bankk}.

 One simple  vacuum  \ishi\  is  
 represented by $\X_a=0$ and 
 $\X_0=diag(\td x_0^{(1)}, ...,
  \td x_0^{(N)})$ 
 with  diagonal entries uniformly distributed 
  on an interval $(0,2\pi R)$.\foot{This corresponds to an array 
  of instantons  over a circle $\td x_0$ 
  with a Wilson line $A_0=\X_0$ 
  which after $T$-duality  becomes  a euclidean 
  world line  of  a  0-brane.}  
  In the large $N$ limit one can then use (as in \egu)  the  
  derivative 
  representation  
  $T \X_0 = i{\del\ov \del x_0}$. 
   Writing the instanton action \inst\  in  this vacuum 
  following  \ishi\ 
  one finds 
  the D0-brane action, now  in its full NBI form 
  \zerr.  
  
  Starting with the  action 
  \zerr\  for a large number $N$ of 0-branes 
  and expanding it 
  near configurations with $i T^{-2} \bar SF_{rs}=
   [\X_r,\X_s] = i f_{rs} I$ ($r,s>0$)
  one can  obtain,  as in \bankk,
  the actions  for type  IIA Dp-branes.
  A periodic array of 0-branes on 2-torus with diagonal
   Wilson lines $A_s=TX_s$ ($s=1,2$) is transformed 
   under T-duality into a 2-brane wrapped over the dual 2-torus.
   On the other hand, the  0-brane configuration
    with $\bar F_{12}=F \not=0$ corresponds to a bound state 
    of a 2-brane with  (a large number $F\inv \sim N$ of) 
    0-branes \refs{\ahber,\gilma,\gil}.
     Under $T$-duality  along the two
    directions of the torus
    the 2+0  system \refs{\pol,\liff, \rust,\mye}  
    goes into  a similar 2+0 system on the dual torus
      with the  charges  
    of 2-branes and 0-branes interchanged.
 The 2+0 non-threshold bound state can be described by a 2-brane action with an
 extra  magnetic flux $\bar F_{12}$ providing a  source for the RR vector
 field \doug. Such an action is indeed what one finds 
 if,  following \bankk,  one introduces the dependence on the two 
 extra coordinates $x_s=(x_1,x_2)$,  sets\foot{Another representation
  used in 
 \refs{\banks,\bankk} is  based on a  
  a pair of non-commutative 
  phase-space coordinates $x_s=(x_0,x_1)$,
 \ $[x_0,x_1]= i  T^{-2} F$, 
   \  with  $\X_0 =x_0, \ \X_1 = x_1$  under the assumption 
    that  fluctuations
of $X_m$   become functions on this phase space (so that 
    $[\X_s, X_a]= i\del_s X_a$, etc.).} 
   \eqn\strii{
 T\bar X_s = i\del_s +  \bar A_s (x)  \ , \ \ \  \ 
  \bar F_{rs}= \del_r \bar A_s -\del_s \bar A_r  = F \ep_{rs} \ ,  }
and substitutes
$X_s = \bar X_s + A_s(x),\  
 X_a= X_a(x)$  into \zerr\ ($A_s,X_a$ become 
 fields of 2-brane action).
 This is effectively a $T$-dual description, in which the flux $F$
 represents the  charge of extra 0-branes on 2-brane, i.e. this
 procedure, just like the $T$-duality relation discussed above, gives
 the `pure' D2-brane action in the limit $F\to 0$.
 A  system  of 0-branes on 4-torus with two fluxes $\bar F_{12}$ and
 $\bar F_{34}$  is $T$-dual to a non-threshold bound state
  `$4+ 2\perp 2+ 0$'  \liff\ 
 (which is  also $T$-dual to a pure 2-brane with the fluxes being 
 related to the two  angles of the 
  $O(2,2)$ duality transformation 
 \mye).
 The  case  with three  fluxes $\bar F_{12}$, $\bar F_{34}$, $\bar F_{56}$
 corresponds to the `$6+4\bot 2+ 2\bot 2\bot 2+0$'  
 configuration  \refs{\bankk,\gil} 
 (which is $O(3,3)$ dual to a pure 2-brane on 6-torus).
 Actions for fluctuations near these solutions 
 obtained from  \zerr\ take the standard D-brane form \qer,\rrer\ 
 in the limit of vanishing background gauge field 
  fluxes.\foot{There is also a 
 `pure 4-brane'  solution $[\X_s,\X_r] = \ha \ep_{srkl} [\X_k,\X_l]$
 \refs{\bankk,\gano}  which should 
 correspond to the threshold $4+0$ 
 bound state in type IIA theory. Setting $T\X_s= i\del_s + \bar A_s$
 ($s=1,2,3,4$) where $\bar A_a$ is an 
 instanton field \refs{\gano,\bankk}  and expanding 
 \zerr\ near such solution 
 one finds the D4-brane action with the vector field 
 $A_s(x) + \bar A_s$.}

 The discussion of  IIB solutions and corresponding Dp-brane actions
  following from \inst\
    is very similar (essentially $T$-dual) to the type IIA one, 
  with the instanton 
 playing the role of the D0-brane, D-string -- the role of  D2-brane,
 D3-brane -- the role of D4-brane, etc. 
 The simplest non-trivial  solution of \eqe\ \refs{\ishi,\lii} 
 is the one with 
   $\bar F_{rs} = -iT^2 [\bar X_r,\bar X_s] =  F \ep_{rs} I $\ 
 for $r,s=0,1$
 and   $\bar X_a=0$ for $a=2,...,9$.
 In complete analogy with the 
  2+0 type IIA configuration mentioned above to which 
  it is $T$-dual, 
  it   describes  a (euclidean) D-string 
   the world sheet  `populated'
  by 
   a large number of instantons,  a non-threshold type IIB 
    bound state 
   which we shall denote as  $1 +i$
   (the $F_{01}$ flux provides a source for the instanton $C$-field).   Our interpretation 
   of the string solution is   thus 
    somewhat different from that of \refs{\ishi,\lii}
   where this configuration was considered as  
    a pure D-string. 
   The presence of 0-branes on type IIA 
   D2-brane in \refs{\bankk,\gilma}  (and the absence 
   of  anti-0-branes in 
    the matrix model of \banks) 
   is directly related to  the basic fact  
   that  the corresponding 
   M2-brane is boosted in 11-th dimension
   (transversely boosted M2-brane has 2+0 configuration as its
   type IIA counterpart 
    \rust). 
    Given that  the D-instanton 
   charge can be interpreted as a momentum in 12-th dimension
    \tsey, 
   this suggests, by analogy, 
    that the instanton matrix model of  \ishi\ or \inst\
   may, in some sense, correspond
     to an `infinite momentum frame' in
   a hypothetic  $D=12$ theory. 
    A reflection of this is
   that all type IIB Dp-branes  appear in 
   this model  as  being bound 
    to a large number of instantons.

 Representing the solitonic
  string  background as in \strii\ 
  ($ \bar A_r(x)  = - \ha F \ep_{rs}  x_s$)
 \eqn\moo{
 \bar X_0 = -\m x_1 + i T\inv \del_0\  , \  \ \ \ \ \  
 \bar  X_1 = \m x_0 + iT\inv \del_1 \ , \ \ \ \ \ \
  \m\equiv  -\ha FT\inv \  ,   }
 and plugging $X_s = \bar X_s +
 A_s(x),\  
 X_a= X_a(x)$ into \inst\ (with  the fluctuations  
 $A_s, X_a $  being  abelian  
  in the  single string case, and  $\STr \to \int d^2x $)
 one finds 
 the   $p=1$  action \qer\ 
   with the 2-d gauge field  $A_s \to   \bar A_s + A_s(x)$. For
   $F=0$ or $\bar A_s=0$, 
    or,  in the $T$-dual picture,  for the 
     zero instanton charge on the string, 
     the resulting   action 
   becomes  the standard  Born-Infeld  (or static gauge Nambu) 
    D-string action.\foot{Having the NBI action 
 \inst\ as a starting point  makes unnecessary to  use
 the  constructions \refs{\ishi,\lii} 
   with an extra `chemical potential'
   term and an independent `area' variable 
     in the action in order 
   to establish the
 relation between the instanton matrix model 
 action  and the   Nambu 
 string action.}

 Solutions with fluxes 
   $[\X_r,\X_s]= i f_{rs} I$  in two or three orthogonal 
   planes correspond to 1/2 supersymmetric 
   non-threshold bound states 
   of type IIB  D-branes, namely, 
   `$3+1+i$' and 
   `$5+ 3 + 1 +i$' 
   (these  can also be represented as $O(d,d)$ duals of 
   D-strings or $1+i$ on the corresponding tori).\foot{The 
   self-dual
   solution \refs{\gano,\bankk} 
    $[\X_s,\X_r] = \ha \ep_{srkl} [\X_k,\X_l]$
    here represents the threshold bound state 
     of a D3-brane 
    and instantons `$3+i$' ($T$-dual to $4+0$).
     The instanton charge is generated by the YM
    instanton on 3-brane world volume \doug\ 
    (there also exists the  corresponding 1/4 supersymmetric type IIB supergravity
    background  \tsey).}
 Expanding \inst\  near these  
 solutions represented as in  \strii\  one finds the 
  D3-brane and D5-brane  BI actions with extra 
  constant  gauge field
 fluxes.

 The above discussion has a  generalisation \bankk\ to the 
   multi-brane solution cases  
 where $\bar X_a$  are  non-vanishing  diagonal matrices
 of Dp-brane  positions and 
  $A_s$ and $X_a$  are  $U(n)$  matrices, 
    so that  one is led to  
   the non-abelian 
 Dp-brane  actions in 
 \rrer.

 To summarise, starting with the non-abelian  
  action \inst\  for  a large number 
 of instantons 
 `filling' some $p+1$  space-time dimensions we  get 
 the   Dp-brane actions   in their   full  
 {\it non-linear}  Born-Infeld form. 
 In a sense, the  world-volumes of all higher branes are thus 
represented by  arrays of instantons  (or `events'), just like 
 type IIA D-branes can be 
   `built' out  of 0-branes.  
At a perturbative $D=10$ type II theory 
level, these relations 
can be viewed as   consequences  of T-duality
implemented   
in the context of large $N$ 
matrix model approaches  
\refs{\banks,\ishi,\bankk}.

Finally, let us recall that (i)
the standard Nambu string action is 
 quadratic in the light-cone gauge but
has  a non-polynomial BI  structure in the static gauge,  
(ii) there is a close 3-d duality relation  between the 
 Nambu-type  $D=11$ 2-brane action and
D2-brane BI action \refs{\dull,\towns,\schm}, and 
(iii) as discussed above, there are  correspondences  between the 
instanton action \inst\ and the static gauge BI D-string action, 
and the  D0-brane action and
D2-brane  static gauge BI actions \rrer.  
These facts put together  may be   suggesting   
 that (a supersymmetric
extension of)  the  
non-linear NBI 
 0-brane  action  \zerr\ may have something to do with  
a { covariant}  formulation
of the  M-theory matrix model of \banks.

\bigskip

I would like to thank  C. Callan, I. Klebanov,  H. Leutwyler, J. 
 Maldacena,   
R. Metsaev  and A. Schwimmer 
for useful discussions at different stages of this work. 
I  acknowledge also  the support of PPARC and  the European
Commission TMR programme grant ERBFMRX-CT96-0045.

\bigskip

\vfill\eject
\listrefs
\end